\newif\ifAnon\Anonfalse
\newif\ifNotes\Notesfalse
\pdfoutput=1
\PassOptionsToPackage{table}{xcolor}

\newcommand{\swallow}[1]{}

\documentclass[sigconf]{acmart}
\setcopyright{acmcopyright}
\copyrightyear{2023}
\acmYear{2023}
\acmDOI{XXXXXXX.XXXXXXX}
\acmConference[CCS '23]{2023 ACM SIGSAC Conference on Computer and Communications Security}{November 26--30, 2023}{Copenhagen, Denmark}
\acmPrice{15.00}
\acmISBN{978-1-4503-XXXX-X/18/06}

\usepackage{array}
\usepackage{multirow}
\usepackage[inline]{enumitem}
\usepackage{moreenum}
\usepackage[utf8]{inputenc}
\usepackage{balance}
\usepackage{cleveref}
\usepackage{footnote}
\usepackage{tablefootnote}
\usepackage[htt]{hyphenat}
\usepackage[english]{babel}
\usepackage[english]{isodate}
\usepackage{oplotsymbl}
\usepackage{subcaption}
\usepackage[load-configurations=abbreviations,binary-units]{siunitx}
\usepackage{booktabs}
\usepackage{xspace}
\usepackage{url}
\usepackage{pgfplots}
\usepgfplotslibrary{statistics}
\pgfplotsset{compat=1.16}
\usepackage{subfiles}
\usepackage{tikz}
\usepackage{titlesec}
\newcommand{\addperiod}[1]{#1.}
\titleformat{\paragraph}[runin]{\bfseries}{\theparagraph}{}{\addperiod}
\titlespacing*{\paragraph}{0pt}{1ex plus 1ex minus .2ex}{{\the\fontdimen2\font}}
\usetikzlibrary{decorations.pathreplacing,positioning, arrows.meta}
\usepackage{listings}
\usepackage{stfloats}
\usepackage{wasysym}
\usepackage{comment}
\usepackage{color}
\usepackage{fontawesome5} %
\usepackage{catchfile}
\usepackage{threeparttable}

\newlist{paraenum}{enumerate*}{1}
\setlist[paraenum]{label=\emph{(\arabic*)}}

\usepackage{array}
\newcolumntype{L}[1]{>{\raggedright\let\newline\\\arraybackslash\hspace{0pt}}m{#1}}
\newcolumntype{C}[1]{>{\centering\let\newline\\\arraybackslash\hspace{0pt}}m{#1}}
\newcolumntype{R}[1]{>{\raggedleft\let\newline\\\arraybackslash\hspace{0pt}}m{#1}}

\usepackage{threeparttable}

\usepackage{pifont}%
\newcommand{\yesmark}{\checkmark}
\newcommand{\nomark}{ }

\newcommand{\etal}{et~al.\ }
\newcommand{\ie}{\textit{i.e.},\ }
\newcommand{\eg}{e.g.,\ }
    
\newcommand{\wrt}{w.r.t.\ }

\newcommand{\binsec}{\textsc{Binsec/Rel}\xspace}    
\newcommand{\binsectwo}{\textsc{Binsec/Rel2}\xspace}
\newcommand{\ctgrind}{ctgrind\xspace}
\newcommand{\dudect}{dudect\xspace}
\newcommand{\abacus}{Abacus\xspace}
\newcommand{\timeout}{\faHourglassEnd}
\newcommand{\true}{\CIRCLE\xspace}
\newcommand{\false}{\Circle\xspace}
\newcommand{\unknown}{\LEFTcircle\xspace}
\newcommand{\crash}{\faBomb\xspace}
\newcommand{\PrimeProbe}{\textsc{Prime+Probe}\xspace}
\newcommand{\FlushReload}{\textsc{Flush+Reload}\xspace}

\newcommand{\nbframeworks}{34\xspace}
\newcommand{\nbprimitives}{25\xspace}
\newcommand{\nblibraries}{3\xspace}

\begin{document}

\title{A Systematic Evaluation of Automated Tools for Side-Channel Vulnerabilities Detection in Cryptographic Libraries}

\ifAnon
\author{}
\else
\author{Antoine Geimer}
\affiliation{%
  \institution{Univ. Lille, CNRS, Inria \\ Univ. Rennes, CNRS, IRISA}
  \streetaddress{UMR 9189 CRIStAL}
  \city{Lille}
  \country{France}
}
\author{Mathéo Vergnolle}
\affiliation{%
  \institution{Université Paris-Saclay, CEA, List}
  \city{Gif-sur-Yvettes}
  \country{France}
}
\author{Frédéric Recoules}
\affiliation{%
  \institution{Université Paris-Saclay, CEA, List}
  \city{Gif-sur-Yvettes}
  \country{France}
}
\author{Lesly-Ann Daniel}
\affiliation{%
  \institution{imec-DistriNet, KU Leuven}
  \city{Leuven}
  \country{Belgium}
}
\author{Sébastien Bardin}
\affiliation{%
  \institution{Université Paris-Saclay, CEA, List}
  \city{Gif-sur-Yvettes}
  \country{France}
}
\author{Clémentine Maurice}
\orcid{0000-0002-8896-9494}
\affiliation{%
  \institution{Univ. Lille, CNRS, Inria}
  \streetaddress{UMR 9189 CRIStAL}
  \city{Lille}
  \country{France}
}

\renewcommand{\shortauthors}{Geimer et al.}
\fi

\begin{abstract}
  To protect cryptographic implementations from side-channel vulnerabilities, developers must adopt constant-time programming practices.
  As these can be error-prone, many side-channel detection tools have been proposed.
  Despite this, such vulnerabilities are still manually found in cryptographic libraries.
  While a recent paper by Jancar \etal shows that developers rarely perform side-channel detection, it is unclear if existing detection tools could have found these vulnerabilities in the first place.
  
  To answer this question we surveyed the literature to build a classification of \nbframeworks side-channel detection frameworks.
  The classification we offer compares multiple criteria, including the methods used, the scalability of the analysis or the threat model considered.
  We then built a unified common benchmark of representative cryptographic operations on a selection of 5 promising detection tools.
  This benchmark allows us to better compare the capabilities of each tool, and the scalability of their analysis.
  Additionally, we offer a classification of recently published side-channel vulnerabilities.
  We then test each of the selected tools on benchmarks reproducing a subset of these vulnerabilities as well as the context in which they appear.
  We find that existing tools can struggle to find vulnerabilities for a variety of reasons, mainly the lack of support for SIMD instructions, implicit flows, and internal secret generation.
  Based on our findings, we develop a set of recommendations for the research community and cryptographic library developers, with the goal to improve the effectiveness of side-channel detection tools.
\end{abstract}

\maketitle

\section{Introduction}

Implementing cryptographic algorithms is an arduous task. Beyond functional correctness, the developers must also ensure that their code does not leak potentially secret information through side channels.
Since Paul Kocher's seminal work~\cite{Kocher96}, the research community has combed through software and hardware to find vectors allowing for side-channel attacks, from execution time to electromagnetic emissions.
The unifying principle behind this class of attacks is that they do not exploit the algorithm \textit{specification} but rather \textit{physical characteristics} of its execution.
Among the aforementioned attack vectors, the processor microarchitecture is of particular interest, as it is a shared resource between multiple programs.
By observing the target execution through microarchitectural components (\eg{} cache~\cite{YaromF14,LiuYG15},  branch predictor~\cite{AciicmezKS07,EvtyushkinRAP18},  DRAM~\cite{PesslGMSM16}, CPU ports~\cite{AldayaBu19}), an attacker can deduce secret information beyond what is normally possible with classical cryptanalysis.
Side-channel primitives using these components allow an attacker to reconstruct secret-dependent control flow and table look-ups.
This problem is exacerbated in multi-core processors and VM environments, where execution can be shared concurrently by multiple actors, and in trusted execution environments, where the privileged control of untrusted operating systems can be leveraged to perform controlled-channel attacks~\cite{DBLP:conf/sp/XuCP15}.

Consequently, multiple countermeasures to timing and microarchitectural attacks have been developed in the literature.
System-level approaches like \textsc{StealthMem}~\cite{KimPM12} modify the OS' behavior between context switches to minimize information sharing, while language-level approaches like type systems can be used to enforce proper information flow in the source code~\cite{AlmeidaBB17}.
Hardware-based approaches also enable securing components by design~\cite{PurnalGG21,ZhangLKE22}.
However, these approaches are  hardly practical as they rely on either large source code rewrites, or require system or hardware modifications.
Manually writing a program such that it is free of such microarchitectural leakage is thus, by far, the most commonly employed countermeasure in cryptographic libraries~\cite{BearSSLConstantTimeCrypto,BernsteinLS12}.
In particular, the \textit{constant-time programming} discipline~\cite{BernsteinLS12,BartheBC14} (CT) consists in writing a program such that its control flow, memory accesses and operands of variable-time instructions do not depend on secret data, and is considered the de-facto standard to protect against (non-transient) timing and microarchitectural attacks. %
Constant-time programming is an arduous task as its recommendations go against usual programming practices and requires knowledge of the literature on side channels.
Moreover, the programmer must be mindful of compiler optimizations not preserving constant-time code~\cite{DBLP:conf/eurosp/SimonCA18,daniel2022binsec} and libraries not following these practices at all~\cite{KaufmannPV16}.

\paragraph{Problem} In the past decade, the research community has investigated \textit{automated} ways of checking whether a program is leakage-free or not.
Many different approaches have been proposed~\cite{DanielBR20,BaoWL21,ctgrind,ReparazBV17,WichelmannME18,WeiserZS18,GrasGK20}, both static and dynamic, but despite this abundance of tools, side-channel vulnerabilities are still found regularly in cryptographic libraries~\cite{LouZJ21}.
Two factors could explain this paradox: either these tools were not able to find such vulnerabilities, or the developers did not use them. %
A recent survey by Jancar \etal\cite{JancarFB22} investigated the opinions of cryptographic libraries developers regarding side-channel detection frameworks. 
They found that while many were willing to include side channels in their threat models, very few actually used tools.
While this provides answers for the second factor, the first one remains unexplored.

\paragraph{Goal} Our paper investigates this question by providing a thorough state of the art on side-channel detection frameworks and recent side-channel vulnerabilities.
In particular, we restrict ourselves to vulnerabilities found in cryptographic libraries spanning five years (2017-2022). %
We only consider vulnerabilities exploited through passive microarchitectural side-channel attacks. Physical side channels~\cite{KocherJJ99, DBLP:journals/iacr/RaoR01, DBLP:conf/ccs/CamuratiPMHF18, DBLP:conf/crypto/GenkinST14}, active attacks (\eg Rowhammer~\cite{KimDKFLLWLM14}) or transient execution attacks~\cite{DBLP:conf/uss/CanellaB0LBOPEG19, DBLP:conf/uss/RagabBBG21, CanellaGGGLMMP019,BulckM0LMGYSGP20} %
are out-of-scope.

We tackle the following research questions: 
\begin{description}
\item[RQ1] How to compare these frameworks, as their respective publications offer differing evaluation? 
\item[RQ2] Could an existing framework have detected these vulnerabilities found manually?
\item[RQ3] What features might be missing from existing frameworks to find these vulnerabilities?
\end{description}

\paragraph{Contributions} The contributions can be summarized as follows:
\begin{enumerate}
\item We present a qualitative classification of the state-of-the-art tools for side-channel vulnerability detection. We classify \nbframeworks frameworks depending on multiple parameters such as the methods used, the type of outputs given by the analysis, or the type of programs analyzed (\Cref{sec:classif-frameworks});   %

\item We compare a subset of these detection frameworks on a unified benchmark, comprised of representative cryptographic operations from \nblibraries libraries, totaling \nbprimitives primitives. We found that asymmetric cryptographic primitives  are still a challenge for most detection tools. This benchmark aims at ensuring a fair comparison between frameworks and help develop  future efforts (\Cref{sec:benchmark});  %

\item We offer a classification of recently published side-channel vulnerabilities %
in cryptographic libraries, offering new insights into where to find potential vulnerabilities (\Cref{sec:attacks});
  
\item We verify whether 4 of these vulnerabilities could have been detected with the aforementioned frameworks. We conclude that support for SIMD instructions, implicit flows, and internal secret generation is crucial for effective side-channel detection tools (\Cref{sec:case-study}). %
\end{enumerate}

\emph{Our complete evaluation framework and results are open source~\cite{web:framework}.}

\section{Preliminaries}\label{sec:preliminaries}

\subsection{Scope}\label{sec:scope}
Language-based approaches for high-assurance cryptography\textemdash in particular
used in the EverCrypt
library~\cite{DBLP:conf/sp/ProtzenkoPFHPBB20,DBLP:conf/ccs/ZinzindohoueBPB17},
FaCT~\cite{DBLP:conf/pldi/CauligiSJBWRGBJ19}, and
Jasmin~\cite{AlmeidaBB17,DBLP:conf/sp/AlmeidaBBGKL0S20}\textemdash provide
provably functionally correct and constant-time implementations of cryptographic
primitives and libraries, but require a complete code rewrite in a dedicated
language. In this paper, we consider out-of-scope such language-based approaches
and focus on tools that are applicable to off-the-shelf libraries.

Program transformations and repair tools have been proposed to transform
insecure programs into (variants of) constant-time
programs~\cite{Agat00,DBLP:journals/ijisec/KopfM07,DBLP:conf/icisc/MolnarPSW05,RaneLT15,DBLP:journals/tcad/ChattopadhyayR18,WuGS18,BrotzmanLZ19,BorrelloDQ21,DBLP:conf/cgo/SoaresP21}.
Because side-channel \emph{detection} is not the main focus of these works, we
only include them when they propose a novel detection phase.

Power side-channel attacks~\cite{KocherJJ99} exploit data-dependent differences
in the power consumption of a CPU to extract secrets. %
These attacks usually assume a stronger attacker model than timing
  microarchitectural attacks, typically requiring physical access to a device. The
  tools we consider in this survey explicitly restrict to the latter attacker
  model, hence they cannot find vulnerabilities related to power-based attacks.

Interestingly, recent \emph{frequency throttling} side-channel
attacks~\cite{WangSP22,DBLP:conf/ccs/LiuCCR22} exploit the fact that
power consumption can also influence the execution time of a program via dynamic
voltage and frequency scaling features of CPUs. These attacks effectively bring
power side-channels to the (micro-)architecture, making them exploitable without
physical access, and blur the boundary between these two attacker models.
We consider these attacks out-of-scope as they also affect constant-time
programs and require different mitigations, traditionally used to protect
cryptographic code against power-side channel attacks, such as
masking~\cite{DBLP:conf/ches/GolicT02,DBLP:conf/fse/Messerges00} or
blinding~\cite{DBLP:conf/ches/JoyeT01}. %

Finally, we focus on side-channel detection tools for
\emph{cryptographic code} and exclude tools targeting other libraries~\cite{Yuan22,WangZL19}.

\subsection{Related Work}

The closest related work is a recent survey by Jancar \etal\cite{JancarFB22,web:ct-tools},
which analyzes how cryptographic developers ensure that their code is secure
against microarchitectural attacks. %
In particular, this survey classifies existing tools for microarchitectural
side-channel analysis, %
and identifies:
\begin{paraenum}
  \item whether developers are aware of these tools or have experience with them,
  \item what kind of tools developers are willing to use (specifically given the trade-off between strong guarantees and usability),
  \item what are the common shortcomings hindering the adoption of these tools.
\end{paraenum}
In contrast, we focus on technical differences between these tools in order to compare their actual capabilities.
We also select a subset of promising tools and experimentally assess their scalability and ability to detect recent vulnerabilities from cryptographic libraries.

Yuan \etal\cite{Yuan22} compared 11 side-channels detection tools on 8 criteria,
  including some overlapping with ours. %
  However, our classification aims at being more general and systematic by
  considering more general criteria and a broader set of tools.

Lou \etal presented a survey~\cite{LouZJ21} characterizing hardware and software vectors in microarchitectural side-channel attacks. While their work is more exhaustive in terms of characterizing side-channel vulnerabilities in cryptographic libraries, our work is focused on vulnerabilities found in a recent period and its purpose is to understand why they were not found automatically.

Barbosa \etal~\cite{BarbosaBB21} proposed a systematization of the literature
and a taxonomy of tools for computer-aided cryptography. Their survey
encompasses formal, machine-checkable approaches for design-level security,
functional correctness, and side-channel security. In contrast, we focus
on side-channel security but consider a broader set of tools, not restricted to
provably secure approaches. %

Complementary to our work, Ge \etal\cite{DBLP:journals/jce/GeYCH18} proposed a
taxonomy of timing microarchitectural attacks and defenses. They give an
overview of microarchitectural components that are susceptible to side-channel
attacks and classify attacks according to the degree of hardware sharing and
concurrency involved. Finally, they survey existing hardware and software
countermeasures against these attacks. Similarly, Jakub
Szefer~\cite{DBLP:journals/jhss/Szefer19} proposed a survey of
microarchitectural channels, attacks and defenses with a stronger focus on
microarchitectural features enabling covert and side channels.

Buhan \etal\cite{BuhanBY21} presented a taxonomy of tools for
protecting against physical side-channel attacks such as power consumption or
electromagnetic emanations.
Their survey encompasses tools for physical side-channel leakage detection,
verification, and mitigation for pre- or post-silicon development stage.
Conversely, our work focuses on (non-overlapping) tools for timing and microarchitectural
side-channels detection in software.

\section{Background}\label{sec:background}

This section introduces the background to understand side-channel vulnerabilities, in particular the hardware and software vectors.%

\subsection{Microarchitectural side channels}

Microarchitectural attacks use shared microarchitectural components to deduce secret information from a program executing on the same physical core, CPU, or socket. %
By probing a component state, the attacker can observe the changes the victim's execution induced on that state, thus gaining information on the victim.
The type and quantity of information gained depend on the component and the way it was probed, e.g.,
branch predictors can expose the direction of branches taken by a victim~\cite{AciicmezGS07}, 
port contention allows an attacker to deduce which instructions the victim executes~\cite{AldayaBu19}, 
TLB attacks leak the victim's memory accesses at page-level resolution~\cite{GrasRB18}. 
By far the most widely used component in attacks remains the cache. 
Multiple techniques have been developed to attack caches, to retrieve both the
victim's control-flow and memory accesses, such as
\PrimeProbe~\cite{Percival05,LiuYG15} and \FlushReload~\cite{YaromF14}. %
Some of these attacks can also be run remotely~\cite{BrumleyB05,BrumleyT11}.
They can be especially powerful in the context of trusted execution
environment, such as Intel-SGX, where the untrusted operating system can be
leveraged to perform controlled-channel attacks, enabling high-resolution and
low-noise side channels~\cite{DBLP:conf/sp/XuCP15,van2020microarchitectural,DBLP:conf/sosp/BulckPS17}.
A complete overview of microarchitectural side-channel attacks can be found in~\cite{DBLP:journals/jce/GeYCH18,DBLP:journals/jhss/Szefer19}.

\subsection{Side-channel vulnerabilities}

The hardware vector represents only one aspect of side channels. To be exploitable, software must leak sensitive data in some way.
Memory accesses and control flow that depend on the secret are the most commonly identified sources of leakage in cryptographic software.
We detail common side-channel pitfalls encountered in cryptographc implementations. An exhaustive look at vulnerabilities in cryptographic libraries is presented in~\cite{LouZJ21}.

\paragraph{Control-flow} One example of secret-dependent control flow is the \textit{square-and-multiply} implementation of modular exponentiation used in RSA, where the operation sequence depends on the bit value of the secret exponent, leading to timing attacks~\cite{Kocher96}. %
Scalar multiplication, the analogous operation for elliptic curve cryptography, suffers from a similar problem in \textit{double-and-add} implementations.
Variants of these implementations such as the \textit{sliding window} and \textit{wNAF multiplication} algorithms are preferred for their performance, but do not alleviate this vulnerability~\cite{Percival05,BengervS14}.
To mitigate control-flow-based side-channel attacks, developers have resorted to either balancing~\cite{AlbrechtP16} or linearizing~\cite{DBLP:conf/icisc/MolnarPSW05} (\ie eliminating) secret-dependent control-flow.
The former solution, employed for instance in the Montgomery ladder algorithm~\cite{JoyeY03} for scalar multiplication, is particularly challenging to get right as it remains vulnerable to attacks exploiting port-contention, branch predictors, and (instruction and data) cache attacks~\cite{YaromB14}.
The latter is not a fool-proof solution either as it remains vulnerable to attacks exploiting secret-dependent memory accesses~\cite{GrasRB18}.
Modular inversion is another common operation that can be a source of side-channel leakage, in particular through the control flow of greatest common divisor (GCD) algorithms, such as the binary extended Euclidean algorithm (BEEA)~\cite{AciicmezGS07}.
Secret-dependent branches are also at the heart of padding oracles used to mount Bleichenbacher's attack~\cite{Bleichenbacher98} or Lucky 13~\cite{AlFardanP13}.

\paragraph{Memory access} The original AES proposal describes how the cipher can be efficiently implemented using pre-computed tables with an access offset depending on the secret key, making it particularly vulnerable to cache attacks~\cite{Bernstein05}.
The aforementioned \textit{sliding window} and \textit{wNAF multiplication} algorithm also employs tables to store pre-computed values to speed up computations which, when accessed with a value derived from the secret, induce leakage through memory accesses~\cite{LiuYG15,BrumleyH09}.

\paragraph{Operand values} Secret-dependent operand values can also be a source of leakage if they influence the program running time.
Early termination in integer multiplication on ARM has been shown to induce a timing leak~\cite{GrossschadlOP10}, with similar problem being found in x86 floating-point instructions~\cite{AndryscoKM15}.

\section{Classifying side-channel vulnerability detection frameworks}\label{sec:classif-frameworks}

As the concept of microarchitectural attacks has  gathered attention only relatively recently, all the approaches we present here are less than a decade old.
We structure this section in two large categories, static and dynamic, that represents a split in approaches but also in research communities: the first closer to program verification and the second to bug-finding.
However, we strive to compare these approaches using broader parameters.
\Cref{tab:soa} gives a comparison of the \nbframeworks detection frameworks we consider in this survey.

\newcommand{\bin}{\textbf{Binary}}
\newcommand{\ct}{\textbf{CT}}
\begin{table*}[t]
  \caption{Comparison table of vulnerability detection frameworks.}
  \label{tab:soa}
  \centering
  \resizebox{\textwidth}{!}{
    \begin{threeparttable}
      \rowcolors{2}{gray!20}{white}
  \begin{tabular}{l l l l l c c c l c c c c c c}
    \toprule
    Ref & Year & \textbf{Tool} & Type & Methods & Scal. & Policy & Sound & Input & L & W & E & B & Available \\
    \midrule
   ~\cite{ctgrind} & 2010 & ct-grind & Dynamic & Tainting & \CIRCLE & \ct{} & \LEFTcircle & \bin{} & \checkmark & & & & \checkmark \\
   ~\cite{BacelarAlmeidaBP13} & 2013 & Almeida \etal & Static & Deductive verification & \Circle & \ct{} & \CIRCLE & C source & & & & & \\
   ~\cite{DoychevFK13} & 2013 & CacheAudit & Static & Abstract interpretation & \Circle & CO & \LEFTcircle & \bin{} & & & \yesmark & & \checkmark \\ %
   ~\cite{BartheBC14} & 2014 & \textsc{VirtualCert} & Static & Type system & \Circle & \ct{} & \CIRCLE & C source & & & \yesmark & & \checkmark \\
   ~\cite{GrussSM15} & 2015 & Cache Templates & Dynamic & Statistical tests & \Circle & CO & \Circle & \bin{} & \checkmark & & & & \checkmark \\
   ~\cite{AlmeidaBB16} & 2016 & ct-verif & Static & Logical verification & \LEFTcircle & \ct{} & \CIRCLE & LLVM & & & & & \checkmark \\
   ~\cite{RodriguesQA16} & 2016 & FlowTracker & Static & Type system & \LEFTcircle & \ct{} & \CIRCLE & LLVM & \checkmark & & & & \checkmark \\
   ~\cite{DoychevK17} & 2017 & CacheAudit2 & Static & Abstract interpretation & \Circle & \ct{} & \CIRCLE & \bin{} & & & \yesmark & & \\
   ~\cite{BlazyPT17} & 2017 & Blazy \etal & Static & Abstract interpretation & \LEFTcircle & \ct{} & \CIRCLE & C source & & & & & \\
   ~\cite{AntonopoulosGH17} & 2017 & Blazer & Static & Decomposition & \LEFTcircle & CR & \CIRCLE & Java & & \yesmark & & & \\ %
   ~\cite{ChenFD17} & 2017 & Themis & Static & Logical verification & \LEFTcircle & CR & \CIRCLE & Java & \yesmark & \yesmark & & & \\
   ~\cite{WangWL17} & 2017 & CacheD & Dynamic & DSE & \LEFTcircle & CO & \Circle & \bin{} & \yesmark & \yesmark & & & \\
   ~\cite{XiaoLC17} & 2017 & STACCO & Dynamic & Trace diff & \LEFTcircle & CR & \Circle & \bin{} & \checkmark & & & & \checkmark \\
   ~\cite{ReparazBV17} & 2017 & dudect & Dynamic & Statistical tests & \LEFTcircle & CC & \Circle & \bin{} & & & & & \checkmark \\
   ~\cite{SungPW18} & 2018 & CANAL & Static & SE & \Circle & CO & \LEFTcircle & LLVM & & \yesmark & & & \checkmark \\
   ~\cite{DBLP:journals/tcad/ChattopadhyayR18} & 2018 & CacheFix & Static & SE & \LEFTcircle & CO & \LEFTcircle & C & \yesmark & \yesmark & & & \yesmark \\
   ~\cite{DBLP:conf/issta/BrennanSBP18} & 2018 & CoCo-Channel & Static & SE, tainting & \CIRCLE & CR & \LEFTcircle & Java & & \yesmark & & \nomark \\
   ~\cite{DBLP:conf/vstte/AthanasiouCEMST18} & 2018 & SideTrail & Static & Logical verification & \Circle & CR & \CIRCLE & LLVM & \yesmark & \yesmark & \checkmark & & \checkmark \\
   ~\cite{ShinKK18} & 2018 & Shin \etal & Dynamic & Statistical tests & \LEFTcircle & CO & \Circle & \bin{} & \checkmark & & & & \\
   ~\cite{WeiserZS18} & 2018 & DATA & Dynamic & Statistical tests & \LEFTcircle & \ct{} & \Circle & \bin{} & \checkmark & & & \yesmark & \checkmark \\
   ~\cite{WichelmannME18} & 2018 & MicroWalk & Dynamic & MIA & \CIRCLE & \ct{} & \Circle & \bin{} & \yesmark & & \yesmark & & \checkmark \\
   ~\cite{Schaub19} & 2019 & STAnalyzer & Static & Abstract interpretation & \CIRCLE & \ct{} & \CIRCLE & C & \checkmark & & & & \yesmark \\
   ~\cite{NilizadehNP19} & 2019 & \textsc{DifFuzz} & Dynamic & Fuzzing & \LEFTcircle & CR & \Circle & Java & & \yesmark & & & \checkmark \\
   ~\cite{WangBL19} & 2019 & CacheS & Static & Abstract interpretation, SE & \CIRCLE & \ct{} & \Circle & \bin{} & \yesmark & \yesmark & & & \\
   ~\cite{BrotzmanLZ19} & 2019 & CaSym & Static & SE & \LEFTcircle & CO & \CIRCLE & LLVM & \yesmark & \yesmark & \\
   ~\cite{disselkoen2020finding} & 2020 & Pitchfork & Static & SE, tainting & \CIRCLE & \ct{} & \LEFTcircle & LLVM & \yesmark & \yesmark & & & \checkmark \\
   ~\cite{GrasGK20} & 2020 & ABSynthe  & Dynamic & Genetic algorithm, RNN & \LEFTcircle & CR & \Circle & C source & \checkmark & & & & \checkmark \\
   ~\cite{HeEC20} & 2020 & ct-fuzz & Dynamic & Fuzzing & \LEFTcircle & \ct{} & \Circle & \bin{} & \yesmark & \yesmark & & & \checkmark \\
   ~\cite{DanielBR20} & 2020 & \binsec & Static & SE & \CIRCLE & \ct{} & \LEFTcircle & \bin{} & \yesmark & \yesmark & & & \checkmark \\
   ~\cite{BaoWL21} & 2021 & Abacus & Dynamic & DSE & \CIRCLE & \ct{} & \LEFTcircle & \bin{} & \yesmark & & \yesmark & & \checkmark \\ %
   ~\cite{JiangBW22} & 2022 & CaType & Dynamic & Type system & \LEFTcircle & CO & \CIRCLE & \bin{} & \checkmark & & & \yesmark & \\
   ~\cite{WichelmannSP22} & 2022 & MicroWalk-CI & Dynamic & MIA & \CIRCLE & \ct{} & \Circle & \bin{}, JS & \yesmark & & \yesmark & & \checkmark \\
   ~\cite{YavuzFF22} & 2022 & ENCIDER & Static & SE & \CIRCLE & \ct{} & \LEFTcircle & LLVM & \yesmark & \yesmark & & & \checkmark \\
   ~\cite{YuanLW23} & 2023 & CacheQL & Dynamic & MIA, NN & \CIRCLE & \ct{} & \Circle & \bin{} & \yesmark & & \yesmark & \checkmark & \checkmark\textsuperscript{$\dagger$}\\
    \bottomrule
  \end{tabular}
  \begin{tablenotes}
  \item (D)SE: (Dynamic) Symbolic Execution, MIA: Mutual Information Analysis, (R)NN: (Recurrent) Neural Network, E: estimation of the number of bits leaked, L: origin of the leakage, W: witness triggering the vulnerability, B: support for blinding. \textsuperscript{$\dagger$} Not available at the time of writing.
  \end{tablenotes}
  \end{threeparttable}}
\end{table*}

\paragraph{Criteria} In addition to comparing frameworks by their type (static or dynamic), we give a description of the precise method employed.
  While our efforts are independent from~\cite{web:ct-tools}, there is a significant overlap between the two classifications. Yet the present one goes into more details, being enriched with additional criteria.
  Similarly to~\cite{JancarFB22,web:ct-tools}, we detail the type of programs supported by the analysis (\eg C, Java, binary code) and whether the approach is supported by a soundness claim.
While analyzing the source code or LLVM is more practical as the program semantics is more easily extracted than in the binary, doing so poses the risk of missing vulnerabilities introduced by the compiler~\cite{KaufmannPV16,DBLP:conf/eurosp/SimonCA18,daniel2022binsec}.
A full soundness claim (\CIRCLE) offers a formal guarantee that the analysis will not, in principle, accept insecure programs and so should not yield any false negatives. 
In practice, this may not be the case because of vulnerabilities outside the analysis' threat model, or because of bugs in the detection tool.
Conversely, partial soundness (\LEFTcircle) claims cover tools that are sound under some restrictions (\eg partial exploration of loops).
In addition, our classification reports the policy enforced by these tools:
(CT) \emph{constant-time} proscribes secret-dependent control flow, memory accesses, and operands of variable time instructions;
(CO) \emph{cache-obliviousness} requires the sequence of cache hits and misses \wrt a cache model to be independent from the secret input;
(CR) \emph{constant-ressource} requires the execution cost (determined by a fixed instruction cost), to be independent from the secret input;
(CC) \emph{constant clock cycles} requires the number of clock cycles to be independent from the secret input.
  Note that CT~\cite{BartheBC14,AlmeidaBB16} is strictly more conservative than CO and CR, and is the only policy that is secure \wrt the attacker scope considered in this paper.

We also detail the type of outputs given by the analysis: estimation on the number of bits leaked (E), origin of the leakage (L), and whether a witness triggering the vulnerability is given (W).
A tool reporting neither these three only reports the presence or absence of vulnerabilities on the target.
We additionally report whether the tool supports blinding~\cite{DBLP:conf/ches/JoyeT01} (B), a defense that introduces additional randomness to computations to hinder inference of secret values.
Finally, we offer a rough estimation of the tool scalability.

\paragraph{Limitation} Unfortunately, quantifying scalability remains challenging due to the absence of a universally applicable metric.
One option would be to report the number of instructions processed by second by a tool. However not all publications provide this information and the concept of \textit{instruction} can vary from one approach to another (LoC, IR or assembly instructions, counting unrolled loops or not, etc.).
Additionally, such metric is specific to a given set of benchmarks and hardware, whereas publications typically feature diverse benchmarks across various setups.
Hence, we instead chose to provide a rough estimation of scalability, based on the claims made in the tools' publications. We differentiate tools able to analyze in a reasonable time: complex asymmetric primitives (\CIRCLE), symmetric primitives (\LEFTcircle), and those struggling to scale even for these (\Circle). %

\subsection{Static analysis}

Static analysis approaches attempt to derive security properties from the program without actually \textit{executing} it, extracting formally defined guarantees on all possible executions through binary or source code analysis.
As a formal exploration of every reachable state is unfeasible, program behavior is often approximated, making them prone to false positives.  
Static approaches were the first to be considered, as side-channel security is closely related to information flow policies~\cite{DenningD77}.

   \subsubsection{Logical reduction}

Non-interference is a \textit{2-safety property} stating that two executions with equivalent public inputs and potentially different secret inputs must result in equivalent public outputs.
This definition covers side channels by considering \textit{resource usage} (\eg address trace) as a public output.
Approaches based on logical reduction to \textit{1-safety} transform the program so that verifying its side-channel security amounts to proving the safety of the transformed program.

Self-composition~\cite{BartheDR11} interleaves two executions of a program $P$ with different sets of secret variables in a single self-composed program $P;P'$. %
Solvers can then be used to verify the non-interference property.
This approach was used by Bacelar Almeida \etal\cite{BacelarAlmeidaBP13} to manually verify limited examples, relying on a large amount of code annotations.
\texttt{ct-verif} instead runs the two copies in lockstep, while checking their assertion-safety~\cite{AlmeidaBB16}.
It is able to verify LLVM programs, leveraging the \textsc{boogie} verifier.
Sidetrail~\cite{DBLP:conf/vstte/AthanasiouCEMST18} reuses this to verify
that secret dependent branches are balanced (assuming a fixed instruction cost and excluding memory access patterns), providing a counter-example when this verification fails.

However, such approaches suffer from an explosion in the size of the program state space.
Blazer~\cite{AntonopoulosGH17} verifies timing-channel security on Java programs by instead decomposing the execution space into a partition on secret-independent branches.
Proving 2-safety is thus reduced to verifying 1-safety on each trace in the partition improving scalability at the cost of precision.
Themis~\cite{ChenFD17} uses static taint analysis to automatically annotate secret-dependent Java code with Hoare logic formulas as pre- and post-conditions. %
An SMT solver then verifies that the post-condition implies %
execution time differences remain bounded by given constant.
Both tools provide a witness triggering the vulnerability otherwise.

	\subsubsection{Type systems}

        Approaches based on verifying type safety of a program differ from language-level countermeasures~\cite{AlmeidaBB17,BondHK17}, as the developer only needs to type the secret values with annotations instead of rewriting the program.
The type system then propagates this throughout the program, similarly to static taint analysis.
Type systems were considered relatively early to verify non-interference properties~\cite{Agat00} and offer good scalability but their imprecision makes them difficult to use in practice.

\textsc{VirtualCert}~\cite{BartheBC14} analyzes a modified CompCert IR where each instruction makes its successors explicit.
The authors define semantics for that representation, building the type system on top of it.
An alias analysis giving a sound over-approximated prediction of targeted memory address is needed to handle pointer arithmetic.
While this approach is more suited to a strict \textit{verification} task, it can also provide a leakage estimate.

FlowTracker~\cite{RodriguesQA16} introduces a novel algorithm to efficiently compute implicit information flows in a program, and uses it to apply a type system verifying constant-time.

	\subsubsection{Abstract interpretation}
	
        As a program semantics is generally too complex to formally verify non-trivial properties, abstract interpretation~\cite{CousotC77}  over-approximates its set of reachable states,
        so that if the approximation is safe, then the program is safe.

CacheAudit~\cite{DoychevFK13} performs a binary-level analysis, quantifying the amount of leakage depending on the cache policy 
by finding the size of the range of a side-channel function. 
This side-channel function is computed through abstract interpretation, and the size of its range determined using counting techniques.
It was later extended to support dynamic memory and threat models allowing byte-level observations~\cite{DoychevK17} and more x86 instructions~\cite{MantelWK17}.

Blazy \etal\cite{BlazyPT17} focus on the source code instead of the binary.
Their tool is integrated into the formally-verified Verasco static analyzer, and uses the CompCert compiler.
The analysis is structured around a tainting semantics that propagates secret information throughout the program. %

STAnalyzer~\cite{Schaub19} uses data-flow analysis to report secret-dependent branches and memory accesses.

CacheS~\cite{WangBL19} uses an hybrid approach between abstract interpretation and symbolic execution.
The abstract domain keeps track of program secrets---with a precise symbolic representation for values in order to confirm leakage---but keeps only a coarse-grain representation of non-secret values.
To improve scalability, CacheS implements a lightweight but unsound memory model.

	\subsubsection{Symbolic execution}

  Symbolic execution~\cite{DBLP:journals/cacm/King76} (SE) denotes approaches
  that verify properties of a program by executing it with symbolic inputs
  instead of concrete ones. Explored execution paths are associated with a
  logical formula: the conjunction of conditionals leading to that path. A
  memory model maps encountered variables onto symbolic expressions derived from
  the symbolic inputs and the concrete constants. A solver is then
  used to check whether a set of concrete values satisfies the generated formulas.
  Recent advances in SMT solvers have made symbolic execution a practical tool
  for program analysis~\cite{DBLP:journals/cacm/CadarS13}. %

CoCo-Channel~\cite{DBLP:conf/issta/BrennanSBP18} identifies secret-dependent
conditions using taint-analysis, constructs symbolic cost expressions for each
path of the program uses SE and reports paths that exhibit
secret-dependent timing behavior. Their cost model
assigns a fixed cost per instruction, excluding  secret-dependent
memory accesses. %

Several works use symbolic execution to derive a symbolic cache model and check that cache behavior does not depend on secrets.
CANAL~\cite{SungPW18} models cache behaviors of programs directly in the LLVM intermediate representation by inserting auxiliary variable and instructions. It then uses KLEE~\cite{DBLP:conf/osdi/CadarDE08} to analyze the program and check that the number of hits does not depend on secrets.
Similarly, CacheFix~\cite{DBLP:journals/tcad/ChattopadhyayR18} uses SE to
derive a symbolic cache model supporting multiple cache policies.
In case of a violation, CacheFix can synthesize a fix by injecting cache
hits/misses in the program. CaSym~\cite{BrotzmanLZ19} follows the same
methodology and, to improve scalability, includes simplifications of the symbolic state and loop transformations, which are sound but might introduce false positives. %

SE suffers from scalability issues when applied to 2-safety properties like constant-time verification. %
Daniel \etal\cite{DanielBR20} adapt its formalism to binary analysis, introducing optimizations to maximize information shared between two executions following a same path.
Their framework \binsec offers a binary-level CT analysis, performing a bounded exploration of reachable states and giving counterexamples for the identified vulnerabilities. 

Pitchfork~\cite{disselkoen2020finding} combines SE and dynamic
taint tracking. It soundly propagates secret taints along all executions paths,
reporting tainted branch conditions or memory addresses. %
Interestingly, Pitchfork can analyze \emph{protocol-level code} by abstracting
away primitives' implementations using function hooks, and analyzing them separately. %

ENCIDER~\cite{YavuzFF22} combines symbolic execution %
with taint analysis to reduce the number 
of solver calls. It also enables  to specify information-flow
function summaries to reduce path explosion.

\subsection{Dynamic analysis}

Dynamic analysis groups approaches that derive security guarantees from execution traces of a target program.
Some form of dynamic binary instrumentation (DBI) is often used to execute the program and gather events of interest, such as memory accesses or jumps.
Dynamic approaches differ in the events collected, and how traces are processed.
They can be grouped depending on whether they reason on a single trace, or compare multiples traces together. 

	\subsubsection{Trace comparison approaches}
	
\paragraph{Statistical tests}
Statistical tests can be used to check if different secrets induce statistically significant differences in recorded traces. 
Cache Template~\cite{GrussSM15} monitors cache activity to detect lines associated with a target event, then finds lines correlated with the event using a similarity measure.
A first pass using page-level observations instead of lines can be used to improve scalability~\cite{SchwarzlKG22}.
Shin \etal\cite{ShinKK18} use K-means clustering to produce two groups of traces for each line.
The confidence in the partition indicates which line is likely to be secret-dependent.
DATA~\cite{WeiserZS18} employs a Kuiper test then a Randomized Dependence Coefficient test to infer linear and non-linear relationships between traces and secrets. This was later extended to support cryptographic nonces as secrets~\cite{WeiserSB20}.

Mutual information (MI) can be used to quantify the information shared between secret values and recorded traces, with a non-zero MI score giving a leakage estimation.
MicroWalk~\cite{WichelmannME18} computes MI scores between input sets and hashed traces, with leakage location pinpointed using finer-grained instruction-level MI scores.
MicroWalk-CI~\cite{WichelmannSP22} optimizes this process by transforming the traces in call trees, and adds support for JavaScript and easy integration in CI, following recommendations from~\cite{JancarFB22}.
CacheQL~\cite{YuanLW23} reformulates MI into conditional probabilities, estimated with neural networks.
Leakage location is estimated by recasting the problem into a cooperative game solved using Shapley values.
Contrary to other tools~\cite{BaoWL21,WichelmannME18}, CacheQL does not assume uniform distribution of the secret, nor deterministic executions traces.

STACCO~\cite{XiaoLC17} targets control-flow vulnerabilities specifically in TLS libraries running on SGX, focusing on oracles attacks~\cite{Bleichenbacher98,AlFardanP13}.
Traces recorded under different TLS packets are represented as sequences of basic blocks and compared using a diff tool.

Instead of recording traces, dudect~\cite{ReparazBV17} records overall clock cycles and compares their distribution with secret inputs divided in two classes (fix-vs-random).
While this approach is simple and lightweight, it gives certainty that an implementation is secure \textit{up to} a number of measurements.
Contrary to other tools relying on an explicit
  leakage model, dudect directly monitors timings. Hence,
  vulnerabilities to other microarchitectural attacks like Hertzbleed might (in
  theory) be detected by dudect. %

\paragraph{Fuzzing}
Fuzzing techniques can be used to find inputs  maximizing coverage and side-channel leakage.
\textsc{DifFuzz}~\cite{NilizadehNP19} combines fuzzing  with self-composition to find side-channels based on instruction count, memory usage and response size in Java programs.
ct-fuzz~\cite{HeEC20} extends this method  to  binary executables and cache leakage.

	\subsubsection{Single trace}
    
Other approaches use only  one trace to perform the analysis, sacrificing coverage for scalability.
 ctgrind~\cite{ctgrind} repurposes the dynamic taint analysis of Valgrind to check CT by declaring secrets as undefined memory. This solution is easy to deploy and reuses familiar tools, but remains imprecise.

ABSynthe~\cite{GrasGK20} identifies secret-dependent branches using dynamic taint analysis. It employs a genetic algorithm to build a sequence of instructions based on interference maps evaluating contention created by each x86 instructions.

More precise approaches use SE to \textit{replay} the trace with the secret as a symbolic value %
and check for CT violation.
CacheD~\cite{WangWL17} applies this approach to memory accesses. %
Abacus~\cite{BaoWL21} extends it to control-flow vulnerabilities, picking random values to check satisfiability instead of using a SMT solver.
It also includes leakage estimation through Monte Carlo simulation.

Finally, CaType~\cite{JiangBW22} uses refinement types (\ie types carrying a predicate restricting their possible values) on a trace to track constant bit values and improve precision. CaType also supports implementations that use blinding. %

\subsection{Insights}

Despite the relative youth of the field, a wide variety of approaches have been proposed. 
While initially more static approaches were proposed, dynamic ones soon followed after 2017.
This might represent a shift in research communities, from a focus on verification to bug-finding and, critically, scalability.
Indeed, dynamic approaches typically scale better than static ones. Yet, this advantage mainly applies to single-trace analysis approaches. For trace comparison-based approaches, the scalability gain is less obvious as recording multiple traces can be time-consuming, particularly for statistical approaches requiring a large number of traces, or for slower algorithms (\eg RSA).
Single trace analyses however suffer from a critical lack of coverage, which could be alleviated through methods like fuzzing.
SE has become a popular approach for both static and dynamic methods, as recent advances in SMT solvers make it practical for side-channel detection.

Both the static and dynamic communities could benefit from integrating approaches from one another.
For example we find that Abacus' optimization of trying random values to satisfy SMT formulas would pair well with \binsec's optimizations sparing UNSAT formulas.

\section{Classifying side-channel vulnerabilities} \label{sec:attacks}

We give here a (non-exhaustive) overview of microarchitectural side-channel vulnerabilities which were subject to publications in security and cryptography conferences in the past five years, many of which were still found manually by researchers.
Interestingly, most of these vulnerabilities are new manifestations of already-known vulnerabilities (\Cref{sec:attacks-known}) and only few of them actually target new primitives or functionalities (\Cref{sec:attacks-new}).

\subsection{Known vulnerabilities}\label{sec:attacks-known}

Known vulnerabilities can resurface for two reasons: when known-vulnerable functions are used in \textit{new contexts}, or in \emph{new libraries}.
In the first case, developers keep vulnerable functions in the code-base for performance reasons, carefully avoiding using them when manipulating secret data. 
This practice leaves the door open to new vulnerabilities in which these known-vulnerable functions (\eg \textit{square-and-multiply}) are used in a new context (\eg key generation). 
In the second case, the lack of developer awareness may prevent  side-channel mitigation transfer from one library to the other. 
This also includes libraries choosing to only partially mitigate the vulnerability,  despite available secure alternatives. %

	\subsubsection{New contexts}

Featuring a decade-old code-base, OpenSSL is particularly susceptible to this kind of vulnerability, as side-channel protection in one module might not be correctly ported to another.
In particular, OpenSSL sets a CT flag on BIGNUMs marking secret data so that they can be manipulated using secure functions.
Recent publications have shown that such \textit{insecure-by-default} approaches are particularly error-prone~\cite{TuveriHG18, WeiserSB18, AldayaGT19, GarciaHT20, DeAlmeidaBragaFS21}.

\paragraph{ECDSA} Despite extensive research on mitigating scalar multiplication in ECDSA, Garcia \etal~\cite{GarciaB17} identified exploitable usage of modular inversion with the vulnerable BEEA~\cite{AciicmezGS07}. The vulnerable code path was taken because of a missing CT flag on the nonce.

\paragraph{SM2} The integration of the SM2 standard into OpenSSL did not inherit from lessons learned in ECDSA~\cite{TuveriHG18}.
In particular, SM2 signature generation directly called the vulnerable wNAF scalar multiplication with no padding done on the scalar, allowing timing attacks~\cite{BengervS14}.
Failure to set the appropriate CT flag also resulted in the modular inversion using BEEA.

\paragraph{Key generation} The RSA key generation process did not use side-channel protections, as single-trace attacks were thought to be impractical.
Weiser \etal\cite{WeiserSB18} proved this assumption false by exploiting secret-dependent control-flow in the BEEA used in key generation.
Key generation was independently investigated through the test methodology Triggerflow~\cite{AldayaGT19,GridinTriggerflow2019}, finding additional issues stemming from the CT flag.

\paragraph{Key parsing} Similar vulnerabilities were found in OpenSSL and MbedTLS' key format handling~\cite{GarciaHT20}, as key format standards leave a lot of flexibility to implementations.
Differences in key format were found to induce different execution paths for a same operation, some calling vulnerable functions.
This was the case for RSA key parsing/validation, and signatures for some elliptic curves. A similar problem was discovered earlier~\cite{AldayaBu19}. 

\paragraph{SRP} The missing CT flag  steers OpenSSL implementation of SRP, a password-authenticated key exchange protocol, to an insecure variant of  modular exponentiation using \textit{square-and-multiply}~\cite{DeAlmeidaBragaFS21}.

\paragraph{PRG} %
Among the standard designs for pseudo-random generators (PRGs), \texttt{CTR\_DRBG} generates a pseudo-random bit sequence using the AES cipher.
Cohney \etal\cite{CohneyKP20} investigated implementations of \texttt{CTR\_DRBG} in multiple libraries, finding that the T-Table variant of AES was used despite its well-known vulnerabilities.

	\subsubsection{New libraries}

As the most popular cryptographic library~\cite{NemecKS17}, OpenSSL has received considerable attention from side-channel researchers.
However, mitigations implemented in OpenSSL are not necessarily propagated to other libraries. %
MbedTLS RSA implementation uses the \textit{sliding-window} method despite its vulnerability to cache attacks~\cite{Percival05,LiuYG15}.
Despite the implementation's attempt to balance branches by calling the same function in both branches of the square-and-multiply, Schwarz \etal\cite{SchwarzWGMM17} successfully exploited secret-dependent data access using a cache attack within an SGX enclave.
A similar attack was also performed on the so-called left-to-right variant from Libgcrypt, featuring exponent blinding~\cite{UenoH23}.
Still in SGX, the secret-dependent branch itself remains vulnerable to branch shadowing~\cite{LeeSGKKP17}, or to attacks based on interactions between interrupts and instruction execution time~\cite{PudduSHC21}. 
Hassan \etal\cite{HassanGD20} illustrate this issue in the NSS library. %
While NSS' ECC code is forked from OpenSSL, 
mitigations such as nonce
padding~\cite{BrumleyT11} are not implemented in NSS.

Some libraries implement \textit{pseudo constant-time} instead of full constant-time  to keep their code base easier to maintain.
Such mitigations seek to address the Lucky 13~\cite{AlFardanP13} attack by adding dummy MAC verification or random delays.
Ronen \etal\cite{RonenPS18} demonstrate that they do not protect against cache attacks. %
Similarly, Ronen \etal\cite{RonenGG19} showed that padding oracles leading to Bleichenbacher-like attacks were still present in libraries such as MbedTLS, s2n, or NSS, due to early termination in padding verification.

\subsection{Vulnerabilities in new functionalities}\label{sec:attacks-new}

\paragraph{Arithmetic functions} Genkin \etal\cite{GenkinVY17} targeted the X25519 key exchange implemented in Libgcrypt, finding that, while scalar multiplication is done using a \textit{branchless} Montgomery ladder, the underlying finite field arithmetic functions were not constant-time.
An early exit in the modular reduction function allowed the authors to mount a cache attack recovering the key.
Aranha \etal\cite{AranhaNT20} discovered a similar issue in OpenSSL  branchless Montgomery Ladder used for ECDSA.
When initializing the algorithm, a conditional swap is done depending on a secret value.
While the swap itself is constant-time, the following finite field multiplication is not. %

\paragraph{Hash-to-element} The Dragonfly handshake used in Wi-Fi authentication converts a shared password to an elliptic curve point, 
by computing the point coordinate  derived from the password and a counter.
When the obtained point is invalid, the counter is incremented and the operation repeated, thus creating a timing channel. 
While the standard suggests fixing the repetitions number by adding dummy ones, Vanhoef and Ronen~\cite{VanhoefR20} found that such mitigations were often incorrectly implemented, if at all, leaving implementations vulnerable to cache attacks~\cite{DeAlmeidaBragaFS20}.

\paragraph{Post-Quantum Cryptography (PQC)} %
BLISS-B signature generation involves sampling a secret polynomial from a Gaussian distribution.
Sampling methods based on pre-computed tables were shown to be vulnerable to cache attacks~\cite{PesslBY17} and branch tracing attacks~\cite{WalletT21}.
A common approach in PQC is to construct a cryptographic scheme secure against chosen plaintext attacks then making it secure against chosen ciphertext attacks using a generic transformation.
Such transformation can lead to key-recovery attacks if %
not constant time~\cite{GuoJN20}.

\subsection{Insights}\label{sec:attacks-insights}

The majority of recent publications reproduce vulnerabilities which have been long known, but in new context and libraries.
  Analysis  should not focus on detecting constant-time violations in their code, but rather detect their incorrect use in the wider code-base.
  Test methodologies like TriggerFlow~\cite{GridinTriggerflow2019} are promising in this regard, and complementing them with fuzzing approaches could allow for wider exploration of cryptographic libraries. 
  New vulnerabilities are not found in usual cryptographic primitives directly, but in newer protocols/schemes, or in lower-level utility functions.
Detection tools thus need to be able to fully analyze programs, including utility functions, and scale to full protocol runs.

\section{Tools considered} \label{sec:tools}

The rest of this paper is dedicated to comparing these approaches and characterizing them with regards to known vulnerabilities.
We restrict ourselves to five tools: Abacus~\cite{BaoWL21}, \binsec~\cite{DanielBR20}, MicroWalk-CI~\cite{WichelmannSP22}, dudect~\cite{ReparazBV17}, and ct-grind~\cite{ctgrind}.
Together, they are representative of the diversity of methods of the literature and used in practice, with a mix of static and dynamic tools.
The first three tools are from academic publications, while the last two are known to be used by developers~\cite{JancarFB22}.
Our selection is based on availability, ability to analyze binary code, scalability (from the claims in their publications), and guarantees provided by the tool.

\subsection{Abacus}
Abacus~\cite{BaoWL21}  %
first obtains  an execution trace  by running the target program with concrete inputs using the binary instrumentation framework Intel Pin~\cite{LukCM05}. %
Then, the SE engine \textit{replays} the trace with the secret data set to symbolic values.
Formulas are generated only for instructions manipulating secret values, otherwise concrete values are used.
A control flow or memory access is represented as a function $f$ of the secret symbolic inputs $k$ and public concrete inputs $m$.
The problem of whether an instruction violates CT is reformulated as whether $\exists k_1,k_2 \in K, f(k_1, m)  \neq f(k_2, m)$, with
$K$ the set of possible secrets. 
Contrary to prior work using DSE~\cite{WangWL17}, Abacus does not use SMT solvers but randomly picks values for $k_1$ and $k_2$, as it is often sufficient to trigger the leakage.

In practice, the developer must indicate which variable is secret either through an annotation or writing a custom Pin tool.
We noted a flaw in that Abacus only registers the first annotation and ignores the subsequent ones.
As such, for our experiments, we wrote a Pin tool letting us mark as secret an arbitrary number of buffers of arbitrary length. 
Finally, Abacus computes an estimation of the number of bits leaked. %

\subsection{\binsec}
\binsec~\cite{DanielBR20} uses SE for  bug-finding and
bounded-verification of constant-time at binary level. It lifts the target
binary to an IR, performs SE along all
program paths, and checks that control-flow and 
 memory accesses do not depend on  secret. %
More precisely, \binsec uses \textit{relational} SE to model two executions of a program
in the same SE instance, with the same public input %
but distinct secret inputs. %
SMT queries allow checking whether  these pairs of path can diverge on control-flow or memory addresses.
If a query is satisfiable,  the program is insecure and the solver returns a counter-example triggering the vulnerability.
Conversely, if all queries are unsatisfiable and the exploration
is exhaustive, the program is secure. %
In practice, these queries are expensive, so \binsec uses optimizations to
spare unsatisfiable queries, making it scale on secure programs.
However, satisfiable queries may remain a bottleneck on insecure programs.

\noindent\textbf{\binsectwo.} 
The \binsec team has recently developed a second version, \binsectwo~\cite{binsecrel2}, with better general performance and architecture support (based on the latest version of the symbolic engine), easier set up (core-dump based initialization), and dedicated optimization for satisfiable queries---based on evaluation over pre-chosen concrete values.

\subsection{MicroWalk-CI}

  MicroWalk~\cite{WichelmannME18} records multiple executions of the target function under different
  inputs using Intel Pin.
  The recorded trace contains branch targets and memory addresses %
  encountered during the execution.
  Mutual information scores are then computed between the leakage trace
  and the input set, giving a quantification of the number of input bits leaked.
  MicroWalk offers various MI score granularity/performance tradeoffs, ranging from whole program MI---giving a coarse leakage quantification---to single instructions---for exact leakage localization.
 MicroWalk-CI~\cite{WichelmannSP22} adds support for Javascript. 
  Special attention has been paid to the framework's usability, with %
  human readable reports compatible with CI features.

\subsection{dudect}

dudect~\cite{ReparazBV17} detects timing %
  leakage through repeated timing measurements and comparison using a statistical test.
The binary is executed under two different classes of secret inputs: one set to constant values and one set to values randomly selected before each measurement.
The execution timings of the target function are then recorded by probing CPU cycle counters. %
Measurements above a certain percentile $p$ are assumed to be noise and are removed.
An online t-test is then applied to determine whether the two distributions are distinguishable, with leakage being reported if the value passes a predefined threshold.
While this approach is simple and easy to deploy, it  only gives guarantees  %
{\it up to a certain number of measurements}. %

\subsection{ctgrind}

ctgrind~\cite{ctgrind} uses Valgrind's  memory error detector, Memcheck, to detect potential side-channel leakages.
As Memcheck  already detects branches and memory accesses computed on uninitialized memory, side-channel vulnerabilities can be found by marking secret variables as
undefined,  
through a specific code annotation.  %
Internally,  %
Memcheck detects errors by shadowing every bit of data manipulated by the program with a \textit{definedness} bit V~\cite{SewardN05}, 
propagated throughout the execution similarly to a taint analysis and checked when  computing an address or a jump.

Applied to side-channel analysis however, this approach yields a considerable number of false positives, as errors unrelated to secret values are also reported.
Still, ctgrind is particularly popular in cryptographic libraries for its simplicity and ease of deployment:  %
ctgrind is used by 5 out of the 6 cryptographic libraries  performing CT tests as reported by~\cite{JancarFB22}. %

\section{Unified benchmark}\label{sec:benchmark}

To fairly compare the scalability and vulnerabilities reported by existing approaches (\textbf{RQ1}), we create a unified benchmark, comprised of representative cryptographic operations from \nblibraries libraries, totaling \nbprimitives benchmarks. 
\Cref{tab:benchmark_unsafe} presents our results.

\newcommand{\tablebenchmark}{AES-CBC-bearssl (T) & 20 & \false & 0.05 & \timeout & 36 & \false & 0.02 & 0.10 & 36 & \false & 3.65 & 36 & \false & 0.16 & 36 & \false & 1.39 & \false & 100.51 \\
AES-CBC-bearssl (BS) & 0 & \true & -- & 16.77 & 0 & \true & -- & 0.31 & 0 & \unknown & 10.69 & 0 & \unknown & 0.17 & 0 & \unknown & 1.55 & \unknown & \timeout \\
AES-GCM-bearssl (T) & 20 & \false & 0.05 & \timeout & 36 & \false & 0.02 & 0.22 & 36 & \false & 6.23 & 36 & \false & 0.18 & 36 & \false & 1.51 & \false & 8.69 \\
AES-GCM-bearssl (BS) & 0 & \true & -- & 53.32 & 0 & \true & -- & 0.48 & 0 & \unknown & 14.79 & 0 & \unknown & 0.17 & 0 & \unknown & 1.73 & \unknown & \timeout \\
AES-CBC-mbedtls (T) & 20 & \false & 0.1 & \timeout & 68 & \false & 0.03 & 0.17 & 68 & \false & 5.73 & 68 & \false & 0.19 & 68 & \false & 1.54 & \false & 0.63 \\
AES-GCM-mbedtls (T) & 20 & \false & 0.27 & \timeout & 84 & \false & 0.05 & 0.4 & 84 & \false & 261.76 & 76 & \false & 0.19 & 71 & \false & 1.90 & \false & 0.89 \\
AES-CBC-openssl (EVP) & 0 & \unknown & -- & \timeout & $0^{\dag\ddag}$ & \unknown & -- & 21.92 & 0 & \unknown & 103.59 & 0 & \unknown & 0.66 & 9 & \false & 5.35 & \unknown & \timeout \\
AES-GCM-openssl (EVP) & 0 & \unknown & -- & \timeout & $0^{\dag\ddag}$ & \unknown & -- & 21.19 & 0 & \unknown & 104.27 & 70 & \false & 0.71 & 8 & \false & 5.66 & \unknown & \timeout \\
AES-CBC-openssl (T) & 20 & \false & 0.05 & \timeout & 36 & \false & 0.02 & 0.16 & 36 & \false & 3.02 & 36 & \false & 0.18 & 20 & \false & 1.38 & \unknown & \timeout \\
AES-CBC-openssl (VP) & $0^\dag$ & \unknown & -- & 0.06 & $0^\ddag$ & \true & -- & 0.59 & 0 & \unknown & 1.01 & 0 & \unknown & 0.19 & 0 & \unknown & 1.68 & \unknown & \timeout \\
PolyChacha-bearssl (CT) & 0 & \true & -- & 9.72 & 0 & \true & -- & 0.18 & 0 & \unknown & 1.3 & 0 & \unknown & 0.16 & 0 & \unknown & 1.43 & \unknown & \timeout \\
PolyChacha-bearssl (SSE2) & $0^\dag$ & \unknown & -- & 0.05 & 0 & \true & -- & 0.11 & 0 & \unknown & 1.17 & 0 & \unknown & 0.15 & 0 & \unknown & 1.34 & \unknown & \timeout \\
PolyChacha-mbedtls & 0 & \true & -- & 18.62 & 0 & \true & -- & 0.49 & 0 & \unknown & 5.02 & 0 & \unknown & 0.18 & 0 & \unknown & 2.95 & \unknown & \timeout \\
PolyChacha-openssl (EVP) & 0 & \unknown & -- & \timeout & $0^\ddag$ & \true & -- & 21.55 & \crash & \unknown & 100.19 & 0 & \unknown & 0.67 & 8 & \false & 5.41 & \unknown & \timeout \\
Chacha20-openssl & 0 & \true & -- & 0.77 & 0 & \true & -- & 0.09 & 0 & \unknown & 3.77 & 0 & \unknown & 0.15 & 0 & \unknown & 1.36 & \unknown & \timeout \\
Poly1305-openssl & 0 & \true & -- & 0.26 & $0^\ddag$ & \true & -- & 0.35 & \crash & \unknown & 1.19 & 0 & \unknown & 0.17 & 0 & \unknown & 1.70 & \unknown & \timeout \\
RSA-bearssl (OAEP) & 0 & \unknown & -- & \timeout & 2 & \false & 0.03 & \timeout & \crash & \unknown & 356.41 & 87 & \false & 0.57 & 0 & \unknown & 146.52 & \unknown & \timeout \\
RSA-mbedtls (PKCS) & 1 & \false & 1.60 & \timeout & $5^\ddag$ & \false & 0.21 & \timeout & \crash & \unknown & 2193.2 & 39 & \false & 0.96 & 137 & \false & 826.23 & \unknown & \timeout \\
RSA-mbedtls (OAEP) & 1 & \false & 1.46 & \timeout & $5^\ddag$ & \false & 0.21 & \timeout & \crash & \unknown & 2203.89 & 48 & \false & 0.98 & 137 & \false & 847.27 & \unknown & \timeout \\
RSA-openssl (PKCS) & 1 & \false & 26.50 & \timeout & $1^\ddag$ & \false & 0.44 & \timeout & 0 & \unknown & 551.72 & 321 & \false & 1.32 & 46 & \false & 52.06 & \false & 618.73 \\
RSA-openssl (OAEP) & 1 & \false & 29.02 & \timeout & $1^\ddag$ & \false & 0.46 & \timeout & \crash & \unknown & 535.91 & 546 & \false & 1.73 & 61 & \false & 59.90 & \false & 771.3 \\
ECDSA-bearssl & 2 & \false & 1.82 & \timeout & 2 & \false & 0.84 & \timeout & 0 & \unknown & 246.62 & 7 & \false & 0.41 & 0 & \unknown & 109.31 & \unknown & \timeout \\
ECDSA-mbedtls & 0 & \unknown & -- & 0.54 & $0^{\dag\ddag}$ & \unknown & -- & 0.19 & 0 & \unknown & 467.11 & 50 & \false & 0.81 & 132 & \false & 314.26 & \false & 224.30 \\
ECDSA-openssl & 0 & \true & -- & 2252.36 & $1^\ddag$ & \false & 51. & \timeout & 0 & \unknown & 202.72 & 14 & \false & 1.22 & 28 & \false & 13.76 & \unknown & \timeout \\
EdDSA-openssl & 0 & \unknown & -- & \timeout & $0^\ddag$ & \true & -- & 25.63 & 0 & \unknown & 59.87 & 0 & \unknown & 1.03 & 8 & \false & 9.86 & \unknown & \timeout \\}
\begin{table*}[t]
\centering
\caption{Vulnerability detection.  \#V: number of reported vulnerabilities, S: status (\true secure; \false insecure; \unknown unknown),  t: time to first bug in seconds, T: analysis time in seconds.}
\label{tab:benchmark_unsafe}
\resizebox{\hsize}{!}{
\begin{threeparttable}
  \rowcolors{2}{gray!20}{white}
\begin{tabular}{l cccc cccc ccc ccc ccc cc}
  \toprule
 \rowcolors{white} %
 & \multicolumn{4}{c}{Binsec/Rel} & \multicolumn{4}{c}{Binsec/Rel2} & \multicolumn{3}{c}{Abacus} & \multicolumn{3}{c}{ctgrind} & \multicolumn{3}{c}{MicroWalk} & \multicolumn{2}{c}{dudect} \\
\cmidrule(lr){2-5}\cmidrule(lr){6-9}\cmidrule(lr){10-12}\cmidrule(lr){13-15}\cmidrule(lr){16-18}\cmidrule(lr){19-20}
Benchmarks & \#V & S & t (s) & T (s) & \#V & S & t (s) & T (s) & \#V & S & T (s) & \#V & S & T (s) & \#V & S & T(s) & S & T (s) \\
    \midrule

AES-CBC-bearssl (T) & 20 & \false & 0.05 & \timeout & 36 & \false & 0.02 & 0.10 & 36 & \false & 3.65 & 36 & \false & 0.16 & 36 & \false & 1.39 & \false & 100.51 \\
AES-CBC-bearssl (BS) & 0 & \true & -- & 16.77 & 0 & \true & -- & 0.31 & 0 & \unknown & 10.69 & 0 & \unknown & 0.17 & 0 & \unknown & 1.55 & \unknown & \timeout \\
AES-GCM-bearssl (T) & 20 & \false & 0.05 & \timeout & 36 & \false & 0.02 & 0.22 & 36 & \false & 6.23 & 36 & \false & 0.18 & 36 & \false & 1.51 & \false & 8.69 \\
AES-GCM-bearssl (BS) & 0 & \true & -- & 53.32 & 0 & \true & -- & 0.48 & 0 & \unknown & 14.79 & 0 & \unknown & 0.17 & 0 & \unknown & 1.73 & \unknown & \timeout \\
AES-CBC-mbedtls (T) & 20 & \false & 0.1 & \timeout & 68 & \false & 0.03 & 0.17 & 68 & \false & 5.73 & 68 & \false & 0.19 & 68 & \false & 1.54 & \false & 0.63 \\
AES-GCM-mbedtls (T) & 20 & \false & 0.27 & \timeout & 84 & \false & 0.05 & 0.4 & 84 & \false & 261.76 & 76 & \false & 0.19 & 71 & \false & 1.90 & \false & 0.89 \\
AES-CBC-openssl (EVP) & 0 & \unknown & -- & \timeout & $0^{\dag\ddag}$ & \unknown & -- & 21.92 & 0 & \unknown & 103.59 & 0 & \unknown & 0.66 & 9 & \false & 5.35 & \unknown & \timeout \\
AES-GCM-openssl (EVP) & 0 & \unknown & -- & \timeout & $0^{\dag\ddag}$ & \unknown & -- & 21.19 & 0 & \unknown & 104.27 & 70 & \false & 0.71 & 8 & \false & 5.66 & \unknown & \timeout \\
AES-CBC-openssl (T) & 20 & \false & 0.05 & \timeout & 36 & \false & 0.02 & 0.16 & 36 & \false & 3.02 & 36 & \false & 0.18 & 20 & \false & 1.38 & \unknown & \timeout \\
AES-CBC-openssl (VP) & $0^\dag$ & \unknown & -- & 0.06 & $0^\ddag$ & \true & -- & 0.59 & 0 & \unknown & 1.01 & 0 & \unknown & 0.19 & 0 & \unknown & 1.68 & \unknown & \timeout \\
PolyChacha-bearssl (CT) & 0 & \true & -- & 9.72 & 0 & \true & -- & 0.18 & 0 & \unknown & 1.3 & 0 & \unknown & 0.16 & 0 & \unknown & 1.43 & \unknown & \timeout \\
PolyChacha-bearssl (SSE2) & $0^\dag$ & \unknown & -- & 0.05 & 0 & \true & -- & 0.11 & 0 & \unknown & 1.17 & 0 & \unknown & 0.15 & 0 & \unknown & 1.34 & \unknown & \timeout \\
PolyChacha-mbedtls & 0 & \true & -- & 18.62 & 0 & \true & -- & 0.49 & 0 & \unknown & 5.02 & 0 & \unknown & 0.18 & 0 & \unknown & 2.95 & \unknown & \timeout \\
PolyChacha-openssl (EVP) & 0 & \unknown & -- & \timeout & $0^\ddag$ & \true & -- & 21.55 & \crash & \unknown & 100.19 & 0 & \unknown & 0.67 & 8 & \false & 5.41 & \unknown & \timeout \\
Chacha20-openssl & 0 & \true & -- & 0.77 & 0 & \true & -- & 0.09 & 0 & \unknown & 3.77 & 0 & \unknown & 0.15 & 0 & \unknown & 1.36 & \unknown & \timeout \\
Poly1305-openssl & 0 & \true & -- & 0.26 & $0^\ddag$ & \true & -- & 0.35 & \crash & \unknown & 1.19 & 0 & \unknown & 0.17 & 0 & \unknown & 1.70 & \unknown & \timeout \\
RSA-bearssl (OAEP) & 0 & \unknown & -- & \timeout & 2 & \false & 0.03 & \timeout & \crash & \unknown & 356.41 & 87 & \false & 0.57 & 0 & \unknown & 146.52 & \unknown & \timeout \\
RSA-mbedtls (PKCS) & 1 & \false & 1.60 & \timeout & $5^\ddag$ & \false & 0.21 & \timeout & \crash & \unknown & 2193.2 & 39 & \false & 0.96 & 137 & \false & 826.23 & \unknown & \timeout \\
RSA-mbedtls (OAEP) & 1 & \false & 1.46 & \timeout & $5^\ddag$ & \false & 0.21 & \timeout & \crash & \unknown & 2203.89 & 48 & \false & 0.98 & 137 & \false & 847.27 & \unknown & \timeout \\
RSA-openssl (PKCS) & 1 & \false & 26.50 & \timeout & $1^\ddag$ & \false & 0.44 & \timeout & 0 & \unknown & 551.72 & 321 & \false & 1.32 & 46 & \false & 52.06 & \false & 618.73 \\
RSA-openssl (OAEP) & 1 & \false & 29.02 & \timeout & $1^\ddag$ & \false & 0.46 & \timeout & \crash & \unknown & 535.91 & 546 & \false & 1.73 & 61 & \false & 59.90 & \false & 771.3 \\
ECDSA-bearssl & 2 & \false & 1.82 & \timeout & 2 & \false & 0.84 & \timeout & 0 & \unknown & 246.62 & 7 & \false & 0.41 & 0 & \unknown & 109.31 & \unknown & \timeout \\
ECDSA-mbedtls & 0 & \unknown & -- & 0.54 & $0^{\dag\ddag}$ & \unknown & -- & 0.19 & 0 & \unknown & 467.11 & 50 & \false & 0.81 & 132 & \false & 314.26 & \false & 224.30 \\
ECDSA-openssl & 0 & \true & -- & 2252.36 & $1^\ddag$ & \false & 51. & \timeout & 0 & \unknown & 202.72 & 14 & \false & 1.22 & 28 & \false & 13.76 & \unknown & \timeout \\
EdDSA-openssl & 0 & \unknown & -- & \timeout & $0^\ddag$ & \true & -- & 25.63 & 0 & \unknown & 59.87 & 0 & \unknown & 1.03 & 8 & \false & 9.86 & \unknown & \timeout \\
\bottomrule
\end{tabular}
\begin{tablenotes}
  \small
\item \crash: the tool crashed. \timeout: the analysis timed-out after \SI{3600}{\second}. \dag: the analysis terminated early (\eg unsupported instructions). \ddag: the analysis starts from a core-dump (\binsectwo only)

\end{tablenotes}
\end{threeparttable}
}
\end{table*}

\subsection{Description and setup}

\paragraph{Libraries} We include OpenSSL~\cite{web:openssl}, as it is one of the most popular cryptographic library~\cite{NemecKS17} and has a long history of side-channel vulnerabilities.
MbedTLS~\cite{web:mbedtls} is another popular library, particularly for embedded targets. Its maintainers consider side-channel attacks in their threat model, though it is a work in progress relevant mainly for new code.
BearSSL~\cite{web:bearssl} is a smaller library with an emphasis on ease of deployment and security.
In particular, the author makes clear claims on which functions are constant-time, which helps us establish a \textit{ground truth} for our benchmark.
For each of these libraries we pick the latest versions at the times of experiment (3.0.5, 3.1 and 0.6 respectively), and compile the libraries as static 32-bits objects, to meet the detection tool requirements.
While we keep library configurations close to their defaults, for MbedTLS, we disable support for VIA PadLock instructions and AES table generation code.

\paragraph{Algorithms} We chose widely-used cryptographic operations from both symmetric and asymmetric schemes.
We target AES encryption, in both CBC and the authenticated  GCM mode, as well as the Poly1305-Chacha20 AEAD scheme.
We target RSA decryption with both  PKCS\#1 v1.5 and OAEP padding scheme.
ECDSA signature generation using  P-256 and EdDSA signature generation using Curve25519 are also included.

\paragraph{Implementations} A library can include different implementations of a single algorithm with differing impacts on side-channel leakage, notably for AES.
For OpenSSL we include the vulnerable T-table implementation and the secure vector permutation (VP) implementation.
For BearSSL, we include its T-table implementation and its constant-time implementation, based on a bit-slicing (BS) approach.
Such implementations also exist for OpenSSL, but not for our 32-bits configuration.
Finally, we add a benchmark using OpenSSL's EVP API, which is the intended entry point for developers %
and dynamically selects the best implementation available. %

\paragraph{Limits} These  common operations are not all  supported in the same way.
BearSSL only supports RSA decryption with OAEP padding, where only OpenSSL supports EdDSA.
While we favor deterministic ECDSA as it derives the secret from the key without any PRNG, OpenSSL does not support it at the time of our experiments~\cite{opensslPR18809}.
We thus use the non-deterministic %
version.

\paragraph{Benchmark design} %
We limit ourselves to simple test harnesses, limiting the number of auxiliary operations before the target cryptographic operation in order to
avoid introducing extra issues in terms of scalability, instruction support, or non-determinism.
Only the input of the target operation are marked as secret. 
For symmetric ciphers we mark the shared key as secret, for RSA the secret key parameters (including the CRT parameters), 
for elliptic curves the secret scalar. 
For each benchmark, the arrays containing secret values are also marked as such at the beginning of the main. 
While ideally we would run all the selected detection tools on the same binary for a given benchmark, the tools have conflicting requirements. As such, for each benchmark we produce both a 32-bits and a 64-bits version, the latter being used only by MicroWalk.

\paragraph{Evaluation protocol}  We compare the numbers of vulnerabilities found by each tool and the running times of the analysis. All experiments are run on a laptop equipped with an i7-8650U processor running Ubuntu 20.04 LTS.
For Abacus, MicroWalk and both versions of \binsec, the vulnerability count is taken directly from the tool output, corresponding to the number of unique leaky instructions.
For ctgrind, some post-processing is needed on the output.

First, ctgrind reports a single leaking instruction multiple times if it is reached in different contexts, so we deduplicate the reported error count.
Second, as ctgrind is based on Valgrind, errors unrelated to constant-time are also be reported. For example, on statically linked programs, Valgrind reports problems from glibc functions~\cite{bugzilla200535}.
To remedy this, we generate a suppression file by first running the program without annotating the secret. As a result, errors independent of the secret are not reported. %
The running time of the analysis is measured as the total time from first invoking the tool to its exit. For \binsec, this includes generation of the coredump when needed. For Abacus, this includes the time needed to record a trace using Pin, as such our running times are generally longer than those in the original publication~\cite{BaoWL21}.
For dudect, computations of the t-test score are done in batches of 5000 measurements, repeated until the analysis either finishes or times out. New inputs are generated for each measurement. %
We additionally thrash the cache by accessing a large array between each run of the targeted algorithm to limit cache side-effects.
MicroWalk is set up to generate 16 traces, as recommended by the authors~\cite{WichelmannSP22}.

\subsection{Results and discussion}

Table~\ref{tab:benchmark_unsafe} reports, for each tool, the running time of the analysis (T) and the number of vulnerabilities (\#V) found when appropriate.
We also report in column S whether the analysis finds the overall program is secure (\true), insecure (\false) or if no statement can be made (\unknown).
As dynamic tools only model a limited number of executions, they can only report \false and \unknown. Static tools can report \true and \false, reporting \unknown when the analysis times out or stops early without finding vulnerabilities.
For \binsec we also report the time required to find a first vulnerability (t). %

\paragraph{Symmetric primitives} For simpler symmetric cryptography such as AES-CBC and authenticated schemes like AES-GCM or Poly1305-Chacha20, tools that report a number of vulnerabilities tend to agree with each other. In general, the tools scale reasonably well on  such primitives.  
Implementations using constant-time techniques such as vector permutations (VP) or bit-slicing (BS) are reported as secure while those based on table lookups (T) yield a consistent number of vulnerabilities.
    dudect can determine that these implementations are  insecure, but the OpenSSL  T-table implementation as it preloads tables in the cache before accessing them. %

While these numbers are consistent, there are two notable exceptions.
\textbf{First}, ctgrind and MicroWalk show less vulnerabilities in MbedTLS AES-GCM implementation than other tools.
For the former, this discrepancy stems from a bug within Valgrind, where upon encountering two consecutive vulnerable memory accesses, the error reporting for one will ``shadow'' the other.
This is relevant in this benchmark as MbedTLS' GHASH function is implemented using 64-bit integers which, in 32-bit, are compiled into two \texttt{mov} instructions. We have not encountered this issue in other benchmarks.
For the latter, this can be explained by the fact that MicroWalk does not report CT violation where the leakage quantified is too small, whereas other tools will report these violations.
\textbf{Second}, in the  OpenSSL's EVP implementation of AES-GCM, ctgrind reports a large number of vulnerabilities while other tools do not. 
For \binsectwo, this is unsurprising as its analysis stops upon reaching unsupported AES-NI instructions used by the EVP implementation.
The vulnerabilities reported by ctgrind are all located in the GHASH implementation, which employs table look-ups to perform multiplications in $GF(2^{128})$, a potential vulnerability identified in~\cite{KasperS09}.

\paragraph{Asymmetric primitives} In the case of more complex asymmetric cryptography, results vary highly between tools.
For Abacus, no vulnerabilities are found when the analysis finishes, however, we observe multiple crashes.
Analyses often time out on both \binsec and \binsectwo, only finding a handful of vulnerabilities.
For OpenSSL and MbedTLS these vulnerabilities are located in BIGNUM conversion functions called before the target cryptographic operation.
Such functions have been found to be vulnerable before~\cite{YuanLW23,WeiserSB20}, we see here that they are also a source of scalability issues.
For dudect, the need to generate inputs for each measurement is a significant bottleneck, in particular for RSA where key generation is a costly operation.
As a result of slower key generation in MbedTLS and BearSSL, the respective RSA benchmarks time out before finishing a single set of measurements.
In the case of ctgrind, analyses finishes very quickly, however the large number of vulnerabilities found makes interpreting these results complicated, with most vulnerabilities being found in BIGNUM functions.
MicroWalk provides similarly high vulnerability counts, though with analysis times in the order of minutes.

\paragraph{Unsupported instructions} Unsupported instructions can have a crucial impact on the analysi' results.
The vulnerable GHASH implementation is rarely reached in practice, as OpenSSL steers the execution to constant-time variants making uses of carry-less multiplication (\texttt{CLMUL}) instructions.
However Valgrind emulates the results of \texttt{cpuid} to only include instruction sets it supports. Its analysis thus does not reach the \texttt{CLMUL} implementation, instead defaulting to a vulnerable function.
The results we obtained with Abacus contrast with the authors' own RSA and ECDSA benchmarks~\cite{BaoWL21}.
While the authors did not disclose which configurations were used to compile the libraries, disassembling their benchmarks shows that, at least for OpenSSL, assembly implementations were disabled.
Such implementations are included in our benchmark, yielding  SIMD instructions unsupported by Abacus.
While in some cases these cause Abacus to crash, it is also possible they lead to under-tainting, explaining the low vulnerability counts. 

\paragraph{False positives}
Both ctgrind and \binsectwo report violations in RSA-bearssl and ECDSA-bearssl,
which is surprising given BearSSL constant-time policies. A closer inspection
reveals that the RSA implementation computes the actual bit length of RSA primes
by counting the number of nonzero bytes, triggering secret-dependent
control-flow, while ECDSA performs an early exit if the key is not well-defined.
While these violations are technically true CT violations, they are not
exploitable vulnerabilities.
MicroWalk on the other hand does not report these CT violations as they do not result in differences in traces for the inputs we used.
In both cases, \binsectwo reports a subset of the vulnerabilities
  reported by ctgrind, which can be explained by a different notion of
  unique vulnerability between both tools. Indeed, \binsectwo only reports a single violation when leaking the same data on different locations,
  whereas \ctgrind reports multiple locations leaking the same data.
For primitives with dynamic implementation selection (\ie EVP and EdDSA-openssl), MicroWalk reports (likely spurious) violations in the initial selection code, not reported by other tools.
Finally, violations are reported in RSA and ECDSA implementations of
  OpenSSL and MbedTLS, however these implementation use blinding, making the
  violations unexploitable in practice.

\paragraph{Leakage quantification} We did not report the numbers given by Abacus and MicroWalk, as there is no standard way of computing leakage quantification or determining their severity. Moreover, leakage quantification is far from a practical notion of exploitability, as even a single bit of leakage or less can be exploited~\cite{AranhaNT20}.

\paragraph{Improvements over \binsec} %
\binsec may struggles  to fully analyze insecure programs, even simpler ones like AES.
This issue is alleviated in \binsectwo thanks to the chosen value optimization. 
\binsectwo is able to finish analyzing symmetric primitives in times comparable to even dynamic tainting approaches like ctgrind, finding the same number of vulnerabilities. %
While scalability remains an issue for asymmetric primitives,  \binsectwo is able to analyze the OpenSSL implementation of Ed25519 faster than a dynamic method like Abacus. %
The ability to start the analysis from a coredump solves initialization issues (e.g., global function pointers) that cannot be avoided in \binsec, while limiting the amount of instrumentation required.

\section{Case-study: vulnerability validation} \label{sec:case-study}

We now employ the tools described in Section~\ref{sec:tools} to determine if the vulnerabilities in Section~\ref{sec:attacks} could have been discovered automatically (\textbf{RQ2}), and determine features that are missing from existing frameworks to find them (\textbf{RQ3}).

\subsection{Description and setup}

\paragraph{Targeted vulnerabilities} We consider vulnerabilities from three publications:~\cite{AldayaGT19} focused on RSA key generation in OpenSSL 1.0.2k,~\cite{GarciaHT20} focused on  key format handling in OpenSSL 1.1.1a, and MbedTLS 2.18.1 (in particular P256 keys for the former and RSA keys for the latter, and~\cite{GenkinVY17} focused on ECDH decryption in Libgcrypt 1.7.6.
The first two papers describe vulnerabilities stemming from functions known to be vulnerable, but accidentally used in a new context, while the third one represents a new vulnerability. 

\paragraph{Benchmark design} While we are interested in checking for CT violations in these vulnerable functions, a library developer without prior knowledge is unlikely to analyze the right function (\eg GCD, modular exponentiation).
More realistically, such developer would instead check the larger context in which this function is used (\eg RSA key generation).
As such, we are interested in detecting CT violations not just in the vulnerable function itself, but also in its calling context, where detection tools might struggle with scalability, or other usability issues.
We thus run the detection tools on simple programs calling the vulnerable functions directly and the higher-level operation.

\paragraph{Evaluation protocol} We compare the number of vulnerabilities and running times of the analysis from \binsectwo, Abacus, ctgrind and dudect
with their configurations from Section~\ref{sec:benchmark}.

\subsection{Results and discussion}

\newcommand{\tablevalidation}{\textbf{P256 sign (OpenSSL)} &  & 0 & \true & 16.09 &  & \crash & \unknown & 5.69 &  & 0 & \unknown & 0.75 & \checkmark & 33 & \false & 71.44 & \unknown & -- \\
\ wNAF multiplication &  & 5 & \false & \timeout &  & -- & \unknown & \timeout & \checkmark & 222 & \false & 0.54 & \checkmark & 35 & \false & 62.55 & \false & 24.5 \\
\textbf{RSA valid. (MbedTLS)} &  & 5 & \false & \timeout &  & 0 & \unknown & 490.01 & \checkmark & 31 & \false & 0.40 & \checkmark & 41 & \false & 278.94 & \false & 2205.07 \\
\ GCD &  & 5 & \false & \timeout &  & 0 & \unknown & 37.74 &  & 14 & \false & 0.21 & \checkmark & 8 & \false & 22.96 & \false & 4.26 \\
\ modular inversion &  & 3 & \false & \timeout &  & 0 & \unknown & 242.1 & \checkmark & 19 & \false & 0.24 & \checkmark & 17 & \false & 141.82 & \false & 17.7 \\
\textbf{RSA keygen (OpenSSL)} &  & 0 & \true & 0.17 &  & \crash & \unknown & 8.66 &  & 0 & \unknown & 6.36 & \checkmark & 123 & \false & 842.02 & \unknown & -- \\
\ GCD & \checkmark & 3 & \false & \timeout &  & -- & \unknown & \timeout &  & 53 & \false & 0.19 & \checkmark & 4 & \false & 3.61 & \false & 0.7 \\
\ modular exponentiation &  & 1 & \false & \timeout & \checkmark & 4 & \false & 110.73 & \checkmark & 6 & \false & 0.24 & \checkmark & 25 & \false & 13.52 & \false & 6.96 \\
\ modular inversion &  & 1 & \false & \timeout &  & -- & \unknown & \timeout & \checkmark & 115 & \false & 0.21 & \checkmark & 15 & \false & 5.96 & \false & 1.69 \\
\textbf{ECDH decryption (Libgcrypt)} &  & 2 & \false & \timeout &  & 0 & \unknown & 386.76 & \checkmark  & 359 & \false & 1. &  & 34 & \false & 146.50 & \false & 89.46 \\
\ modular reduction &  & 0 & \unknown & \timeout &  & 0 & \unknown & 2.35 & \checkmark & 9 & \false & 0.22 &  & 0 & \unknown & 1.59 & \false & 0.41 \\}
\begin{table*}[t]
\centering
\caption{Vulnerability detection. V: whether the right vulnerability is found \#V: number of reported vulnerabilities, S: status (\true secure; \false insecure; \unknown unknown), T: analysis time in seconds.}
\label{tab:benchmark_validation}
\small
\begin{threeparttable}
\begin{tabular}{l cccc cccc cccc cccc cc}
  \toprule
  & \multicolumn{4}{c}{Binsec/Rel2} & \multicolumn{4}{c}{Abacus} & \multicolumn{4}{c}{ctgrind} &  \multicolumn{4}{c}{MicroWalk} & \multicolumn{2}{c}{dudect} \\
\cmidrule(lr){2-5}\cmidrule(lr){6-9}\cmidrule(lr){10-13}\cmidrule(lr){14-17}\cmidrule(lr){18-19}
Benchmarks & V & \#V & S & T (s) & V & \#V & S & T (s) & V & \#V & S & T (s) & V & \#V & S & T (s) & S & T (s) \\
    \midrule
\textbf{P256 sign (OpenSSL)} &  & 0 & \true & 16.09 &  & \crash & \unknown & 5.69 &  & 0 & \unknown & 0.75 & \checkmark & 33 & \false & 71.44 & \unknown & -- \\
\ wNAF multiplication &  & 5 & \false & \timeout &  & -- & \unknown & \timeout & \checkmark & 222 & \false & 0.54 & \checkmark & 35 & \false & 62.55 & \false & 24.5 \\
\textbf{RSA valid. (MbedTLS)} &  & 5 & \false & \timeout &  & 0 & \unknown & 490.01 & \checkmark & 31 & \false & 0.40 & \checkmark & 41 & \false & 278.94 & \false & 2205.07 \\
\ GCD &  & 5 & \false & \timeout &  & 0 & \unknown & 37.74 &  & 14 & \false & 0.21 & \checkmark & 8 & \false & 22.96 & \false & 4.26 \\
\ modular inversion &  & 3 & \false & \timeout &  & 0 & \unknown & 242.1 & \checkmark & 19 & \false & 0.24 & \checkmark & 17 & \false & 141.82 & \false & 17.7 \\
\textbf{RSA keygen (OpenSSL)} &  & 0 & \true & 0.17 &  & \crash & \unknown & 8.66 &  & 0 & \unknown & 6.36 & \checkmark & 123 & \false & 842.02 & \unknown & -- \\
\ GCD & \checkmark & 3 & \false & \timeout &  & -- & \unknown & \timeout &  & 53 & \false & 0.19 & \checkmark & 4 & \false & 3.61 & \false & 0.7 \\
\ modular exponentiation &  & 1 & \false & \timeout & \checkmark & 4 & \false & 110.73 & \checkmark & 6 & \false & 0.24 & \checkmark & 25 & \false & 13.52 & \false & 6.96 \\
\ modular inversion &  & 1 & \false & \timeout &  & -- & \unknown & \timeout & \checkmark & 115 & \false & 0.21 & \checkmark & 15 & \false & 5.96 & \false & 1.69 \\
\textbf{ECDH decryption (Libgcrypt)} &  & 2 & \false & \timeout &  & 0 & \unknown & 386.76 & \checkmark  & 359 & \false & 1. &  & 34 & \false & 146.50 & \false & 89.46 \\
\ modular reduction &  & 0 & \unknown & \timeout &  & 0 & \unknown & 2.35 & \checkmark & 9 & \false & 0.22 &  & 0 & \unknown & 1.59 & \false & 0.41 \\

\bottomrule
\end{tabular}

\end{threeparttable}
\end{table*}

Table~\ref{tab:benchmark_validation} reports, for each tool, the running time of the analysis (T) and the number of vulnerabilities (\#V) found when available. Abacus report vulnerabilities only upon finishing the analysis, and thus reports none when there is a timeout.
The higher-level operations are listed in bold, while the vulnerable functions used are listed immediately below the corresponding operation.

\paragraph{General results} In general, \binsectwo{}, \ctgrind{}, and \dudect{} report the same programs as secure (\ie{} P256 sign and RSA keygen), and the rest as insecure,
with the exception of the modular reduction in Libgcrypt, for which \binsectwo{} times-out without finding vulnerabilities.
In contrast, \abacus{} only reports a single program as insecure. This is possibly due to unsupported instructions, once again limiting the tool ability to properly taint secret data. Additionally, the analysis will times out in multiple cases, returning no results. The ability to output violations as soon as they are found would, thus, greatly improve usability.
In general, we find that \ctgrind{} is able to find most relevant vulnerabilities in both the target functions and their calling contexts.
While the relevant vulnerabilities can generally be found,
the high number of reported vulnerabilities complexifies the interpretation of these results.
We note that most of these vulnerabilities stem from BIGNUM manipulation functions.
These CT violations are often already known by developers, and so, as pointed out by Jancar \etal\cite{JancarFB22}, the ability to ignore violations in parts of the code would represent a clear usability improvement.
Contrasting with ctgrind, \binsectwo{} found few vulnerabilities.
We note that the analysis often gets stuck early in the program, during BIGNUM conversion steps, and times out. Specifically, benchmarks involving GCD computations and modular inversion are susceptible to path explosion, a common limitation of static SE, which leads to unexplored program behaviors.

\paragraph{Implicit flows} Implicit flows happen when the value of a variable may differ depending on secret-dependent control flow.
None of the four tools we consider %
tracks implicit information flows.
Our benchmarks shows that ignoring them leads to an underestimation of the number vulnerabilities in cryptographic programs.
For example, MbedTLS implements a comparison function in which the return value depends on whether a secret-dependent branch is taken (a typical case of an implicit flow).
MbedTLS GCD function uses the result of this comparison function in a (secret-dependent) conditional jump.
As a result, ctgrind %
only reports the secret-dependent branch in the comparison function as vulnerable, but not the one in the GCD computation. %
While a CT violation is still reported, a developer is not able to fully appreciate how vulnerable the program is without taking into account these implicit flows.

\paragraph{Internal secrets} The way the secret is utilized in some operations can pose problems for detection tools.
For RSA key generation, there is by definition, no initial values to mark as secret.
For P256 signature generation, the secret exploited in publications is the cryptographic nonce generated internally. Other operations making use of PRNG are also affected similarly.
In both of these cases our ``black-box'' experimental setup does not allow us to mark the secret appropriately and as such, the detection tools we used cannot find vulnerabilities.
In practice, it is possible to mark such secrets by deviating from our setup.
For example, the authors of Abacus wrote a special pintool to mark the cryptographic nonce as secret,
but their solution lacks generality. 
Another option is to add annotations within library's source code, something that developers are generally opposed to.

\section{Recommendations}\label{sec:discussion}
We now formulate recommendations to the research community and to cryptographic library developers based on the insights obtained with our experimental evaluation.

\subsection{Recommendations to research community}

 \paragraph{R1. Support of SIMD instructions} Research prototypes like Abacus often only support basic x86 instructions. %
 However, as shown in \Cref{sec:benchmark}, libraries heavily use SIMD instruction sets (\eg AVX2), now commonly supported in CPUs. These implementations
 are now typically selected in priority. 
 As such, supporting SIMD instructions in detection tools is now crucial, and evaluating a tool on implementations specifically selected to \textit{not} use SIMD instructions   %
 may mislead developers on the tool usability. 

 \paragraph{R2. Support of implicit information flows} Future tools should  investigate the possibility of detecting CT violations stemming from implicit information flows.
    We argue that by ignoring them, a detection tool can underestimate the amount of vulnerabilities, giving developers a false sense of security.
    Yet, considering implicit flows is challenging and can impact scalability. %
    Only three tools explicitly support implicit flows~\cite{RodriguesQA16,JiangBW22,YuanLW23}, while trace-comparison-based dynamic tools might only provide a partial support.

    \paragraph{R3. Support of internal secrets} The community should investigate side-channel vulnerabilities in operations where the secret is generated internally, such as
      key generation and PRNG~\cite{AldayaGT19, PudduSHC21, CohneyKP20}. 
    Yet, none of the tools in our classification explicitly support them.
    Recently, a test methodology to detect usage of vulnerable functions in key generation was proposed~\cite{GridinTriggerflow2019}, however it requires knowing vulnerable points in advance.
    
    \paragraph{R4. Support for randomization-based defense}
    Blinding 
    introduces randomization during computations to hinder inference of
    secrets via side-channels~\cite{DBLP:conf/ches/JoyeT01}.
    It poses an additional challenge for detection
    tools as the leakage does depend on the secret, but not in an
    exploitable way, leading to false positives. Only 3 out of the 34 tools 
    we considered in \Cref{tab:soa} support blinding.

 \paragraph{R5. Usage of a standardized benchmark} We  recommend the community to adopt a standardized benchmark of cryptographic implementations to evaluate side-channel detection tools.
    Such benchmark would facilitate comparing detection tools to one another, in particular in terms of usability and scalability.
    We thereby propose to the community our benchmark, which can be readily adopted and extended in the future. 

 \paragraph{R6. Improve usability} %
    As pointed out in \Cref{sec:case-study}, reporting results before the analysis end is crucial for benchmarks that times out.
    We also echo a point made in~\cite{JancarFB22}: being able to ignore vulnerabilities from some parts of the code helps interpreting the results, particularly for approaches like ctgrind.
    For static tools, the ability to write stubs adding limited support for system calls and dynamic allocation was essential in our experiments with \binsec, as
    we cannot expect developers to avoid using them. 
    The ability to start the analysis from a coredump (as in \binsectwo) also greatly simplifies the instrumentation needed to initialize the analysis.

\subsection{Recommendations to developers}
\paragraph{R7. Make libraries more analysis friendly} Static analysis tools often require more instrumentation than dynamic ones. %
In particular, runtime implementation selection (\eg EVP) poses a challenge.
While for experiments using \binsectwo we were able to bypass this issue by initializing the memory from a coredump, we can only analyze one implementation and have little control over which one is chosen.
We recommend that libraries expose in their API, at least for testing purposes, functions to directly call different implementations, as done for example by OpenSSL AES implementations. 

\section{Conclusion}\label{sec:conclusion}

We  surveyed and classified   state-of-the-art side-channel detection tools, and propose a unified benchmark allowing  fair experimental comparison. 
Our findings show that existing tools can still struggle to analyze complex primitives and can miss vulnerabilities.
We issued recommendations for the research and developer community to improve existing tools and encourage their usage.

Beyond these recommendations we note that the research communities for microarchitectural and physical side-channel detection both evolved independently, with little overlap~\cite{BuhanBY21}.
With the advent of frequency scaling side-channels like Hertzbleed~\cite{WangSP22} extracting a timing side-channel from power consumption, this boundary is blurred.
To properly address such vulnerabilities, the two communities should focus on bridging the gap between them, borrowing insights from physical side-channel detection and move beyond the constant-time model.

\ifAnon
\else
\begin{acks}
This work benefited from the support of the ANR-19-CE39-0007 MIAOUS, ANR-20-CE25-0009 TAVA and PEPR PP Secureval projects. %
\end{acks}
\fi

\bibliographystyle{ACM-Reference-Format}
\bibliography{biblio}

\appendix

\end{document}